# Turbulent Convection in Thin Accretion Disks

Itzhak Goldman[1] and Amri Wandel[2]

[1]School of Physics and Astronomy, Sackler faculty of Exact Sciences
Tel Aviv University, Tel Aviv 69978, Israel

[2]Racah Institute of Physics, Hebrew University, Jerusalem, Israel

**Abstract**

A self-consistent solution for a thin accretion disk with turbulent convection is presented. The turbulent convection plays a double role: it provides the disk viscosity and takes part in the vertical transport of the released energy. Rather than assuming arbitrary phenomenological parameterizations for the disk viscosity, the latter is derived from a physical model for turbulence. Employing this model, we express the turbulent viscosity and the vertically averaged convective flux in terms of the local physical conditions of the disk which, in turn, are controlled by the former two. The resulting self-consistent disk structure, and the ratio between the convective and total fluxes are obtained for radiation and gas pressure dominated regions, and for electron scattering and free-free absorption opacities.

In the gas pressure region, two distinct solutions are obtained. In one, the convective flux is much larger than the radiative flux and the blackbody region extends over the entire gas pressure region and could also extend down to the inner boundary of the disk. In this solution the temperature profile is close to adiabatic. In the other solution, the convective flux is about a third of the total flux, the dimensionless superadiabatic temperature gradient is $\sim 0.6$ and there exist the gas pressure blackbody and electron scattering regions as well as the radiation pressure region.

In the radiation pressure region, the temperature profile is very close to adiabatic, and the disk is geometrically thin and optically thick even for super Eddington accretion rates. The fraction of the convective flux, out of the total flux, increases with the accretion rate, and for accretion rates comparable to the Eddington limit is close to 1. This variation stabilizes the, radiation pressure region, so that unlike the $\alpha$ disk, all the disk regions are secularily stable. The values of the effective $\alpha$-parameter are rather small: $\lesssim 5 \times 10^{-4}$, $\sim 1 \times 10^{-3}$ and $\sim 5 \times 10^{-3}$ for radiation pressure region and for the two solutions in the gas pressure region, respectively.

*Subject headings*: accretion, accretion disks — convection — turbulence

## 1. Introduction

There is quite a large body of observational data suggesting the existence of accretion disks around protostars, compact stellar objects and active galactic nuclei (AGNs). No wonder that accretion disk models are extensively employed to interpret these observations. However, to a large extent the models used are more a descriptive than predictive tool (Pringle 1981). The reason for this state of affairs is the lack of a physical model for the disk turbulent viscosity, $\nu_t$, usually parametrized in the form (Shakura & Sunyaev 1973)

$$\nu_t = \alpha c_s h. \tag{1}$$

Here $c_s$ is the speed of sound, $h$ is a characteristic scale height, and $\alpha$ is a dimensionless parameter which neither value nor dependence on the physical conditions in the disk are known. Thus, often, the observations are used to fit the $\alpha$-parameter for a given system. Usually a constant $\alpha$ is assumed for gas pressure dominated regions of the disk. For radiation pressure dominated regions both the former prescription and $\alpha$ proportional to the ratio between the gas and total pressure, have been suggested. In the lack of a physical model for the turbulent viscosity it is not clear which prescription, if any, is the correct one.

A self-consistent solution to the disk equations requires that the turbulent viscosity be determined by the physical conditions of the disk. In turn, the resulting turbulent viscosity will control the same physical conditions. The determination of the turbulent viscosity requires a model for turbulence and the specification of the particular instability that generates the turbulence. The predictions of the self-consistent solution, obtained for the assumed generation mechanism, can then be compared to the observations thus allowing to decide how relevant is the above generation mechanism to the astrophysical system under study.

In spite the very large Reynolds numbers, Keplerian disks are stable against shear-generated turbulence, at least in the linear analysis (Pringle 1981). On the other hand it has been pointed out that they are unstable against turbulent convection, both in radiation pressure and gas pressure regions and for various opacities (e.g., Bisnovatyi-Kogan & Billinikov 1977; Shakura, Sunyaev, & Zilitinkevich 1978; Lin & Paploizou 1980). Therefore, in this paper we implement the above proposed selfconsistent scheme for the case of turbulent convection in a thin accretion disk. The turbulence model employed (Canuto, Goldman & Chasnov 1987) provides vertically averaged values for the turbulent viscosity and for the convective flux. Thus, no attempt to resolve the vertical structure of the disk is made, and all quantities are either vertical averages or midplane values. A detailed vertical structure requires a turbulence model, that could yield the z-dependent values of the turbulent viscosity and of the convective flux. Such a model would be considerably more complex as the effects of the turbulence could no longer be represented by a turbulent viscosity and a convective flux but include also diffusion terms of various turbulent ensemble averages, which vanish in a vertically averaged model. In this work, we wish to focus on the more general features of the proposed self-consistent approach and avoid the complications of a z-dependent model. We note that turbulence bulk properties such as the turbulent viscosity and the convective flux are contributed mainly by the largest eddies. The vertical extent of the latter is $\sim h$ and thus vertical averaging is expected to represent fairly well their contribution. Indeed such averaging was shown to yield quite good predictions regarding laboratory convection (Canuto *et al.* 1987). Thus, our working assumption is that the important characteristics of the self-consistent solution would be evident also in the simpler vertically averaged approach.

Doubts concerning the direction of the angular momentum flux associated with convection have been raised by Ryu & Goodman (1992) and by Kley, Papaloizou, & lin (1993). In the first work linear convectives modes were shown to give rise to an inward angular momentum flux. However, as noted by the authors, this result may just reflect the inconsistency of their model and in a fully developed turbulence the situation would probably be different. Kley *et al.* (1993) assumed an imposed underlying large viscosity and performed direct numerical simulation for convective perturbations. They interpret their results as indicating an inward angular momentum flux advected by convection. As the authors note this may be the result of the large value of the non convective viscosity that was used. Indeed, Cabot & Pollack (1992) find that when the Reynold numbers are increased ( viscosity lowered) turbulent convection provides a positive turbulent viscosity, thus ensuring an outward angular momentum transport. Since in the present model the turbulent viscosity is positive definite (see eqs. [13] and [19]) there will be an outward flux of angular momentum as required.

We have in mind disks surrounding compact objects of sizes ranging from stellar to galactic. For AGN accretion disks, the radiation pressure region can encompass a large part of the disk region from which most of the radiation is emitted. Therefore, we are interested in a self-consistent disk description for the following three regions: radiation pressure dominated region, gas pressure dominated region with electron scattering opacity, and gas pressure dominated region with free-free absorption opacity. Since turbulent convection is taken to be the source of the disk turbulent viscosity, the convective flux could play an important role in the transfer of energy to the disk surface.

The resulting disk structure, and the ratio between the convective and total fluxes at given distance from the central object, are obtained as functions of $\delta-$ the vertically averaged local value of the dimensionless superadiabatic temperature gradient. Practically, it is more convenient to express $\delta$ in terms of $\xi_0-$ the ratio between the convective and total fluxes and use the latter in the parameterization of the disk solutions. Unlike the $\alpha$-parameter, $\delta$ could be determined once the vertical structure of the disk is obtained. Even in the vertically averaged approximation, considered in the present work, we derive general upper bounds on $\delta$. Moreover, using approximate vertically averaged relations we estimate the ratio between the surface and midplane temperatures. This together with an expression for the flux emanating from the disk, in terms of the surface temperature, provides an additional relation that can be used to find $\xi_0$ and $\delta$. Due to the vertical averaging, the obtained solutions are correct up to factors of order unity. Nevertheless, we expect the results to be qualitatively representative of those of a detailed $z$-dependent model.

Our main results are (1) In the radiation pressure dominated region the dimensionless superadiabatic temperature gradient is very small, so that the vertical temperature profile is very close to adiabatic. This result is independent of the ratio of the convective flux to total flux. The ratio of convective to total flux is larger the larger is the accretion rate. For accretion rates comparable to the Eddington limit it varies between $\sim 0.5$ and close to 1. (2) For the gas pressure dominated region we find two distinct types of solutions. In one, practically all the flux is transported by convection and the dimensionless superadiabatic temperature gradient is small ( $\lesssim 0.1$). The blackbody region encompasses the entire gas pressure region and could extend down to a small distance from the central object, without any radiation pressure region. In the other solution, the convective flux comprises about 1/3 of the total flux and the dimensionless superadiabatic gradient is $\lesssim 0.6$. In this case all three regions of the disk exist. 3) The effective $\alpha$-parameter is expressed as a function of the local dimensionless superadiabatic temperature gradient. Its typical values are quite small (this is a general feature of any three dimensional turbulence which has no energy sources external to the disk; Goldman 1991). 4) In the radiation pressure region an increase in the luminosity leads to an increase in the fraction of the convective flux which in turn leads to a net increase in the surface density. Consequently, the surface density is an increasing function of the luminosity, in all disk regions. Therefore, all regions are secularily stable, in contrast to $\alpha$ disks that are unstable in the radiation pressure region.

The role of convection in accretion disks has been discussed by many authors within the phenomenological $\alpha$ -parameter model for the turbulent viscosity, and the phenomenological mixing length approach for the convective flux (see, e.g., Bisnovatyi-Kogan & Billinikov 1977; Livio & Shaviv 1977; Liang 1977; Shakura *et al.* 1978; Vila 1978; Lin & Paploizou 1980; Tayler 1980; Smak 1982; Ruden & Lin 1986; Duschl 1989, Milsom, Chen, & Taam 1994).

Cabot *et al.* (1987a,b) applied an earlier, less developed, version of the turbulence model employed here (Canuto & Goldman 1985) to the protosolar accretion disk. The dominant opacity for that cool disk is due to dust grains. As stated above, we are interested here in much hotter disks surrounding compact objects, with radiation pressure dominated regions and with different opacities. Recently, Cabot *et al.* (1990) and Cabot & Pollack (1992) applied direct numerical simulations to convective disks with uniform and differential rotation, respectively. However, as is the case with direct numerical simulation in other problems involving turbulence, they are limited to low Reynolds numbers which are orders of magnitude smaller than those characterizing astrophysical accretion disks.

Some of the ideas discussed here were presented in an earlier work (Wandel & Goldman 1991).

## 2. Disk Equations

Let us consider a stationary geometrically thin accretion disk in which $h \ll R$, where $h$ is the disk half-width at a distance $R$ from the central object. The radial gradient of the pressure is small compared to the gravitational force per unit volume exerted by the central object, so that the disk angular velocity is Keplerian

$$\Omega = \left(\frac{GM}{R^3}\right)^{1/2} = 2 \times 10^{-3} M_8^{-1} r^{-3/2} s^{-1}, \tag{2}$$

with $M$ the central object mass, $M_8 = M/(10^8 M_\odot)$, and $r = Rc^2/(GM)$.

The vertically averaged disk equations include the energy equation, the angular momentum equation, and the hydrostatic equilibrium equation. The energy released by the turbulent viscosity, per unit area of each face of the disk, $Q$, is given by (Pringle 1981)

$$Q = \frac{9}{4}\nu_t \rho h \Omega^2, \tag{3}$$

where $\rho$ is the density at the disk midplane and $\nu_t$ is the vertically averaged value of the turbulent viscosity. The angular momentum equation expressing outward radial transfer of angular momentum, due to the interaction of the turbulent viscosity with the Keplerian shear, is

$$\Omega \dot{M} \phi(r) = 6\pi \Omega h \rho \nu_t. \tag{4}$$

where $\dot{M}$ is the rate of mass accretion, and $\phi(r)$ accounts for the inner, stress-free, boundary of the disk; for a nonrotating black hole $\phi(r) = 1 - (6/r)^{1/2}$. Combining equations. (3) and (4) results in

$$Q = \frac{3}{8\pi}\Omega^2 \dot{M} \phi(r) = 1.2 \times 10^{20} L_* M_8^{-1} r^{-3} \,\text{erg s}^{-1}, \tag{5}$$

where equation (2) was used to express $\Omega$, and where the accretion rate $\dot{M}$ was expressed in terms of the modified Eddington ratio $L_*$,

$$L_* = \frac{L}{L_E}\phi(r) = 0.4\left(\frac{\dot{M}}{10^{26} g s^{-1}}\right) M_8^{-1} \phi(r). \tag{6}$$

Here $L_E$ is the Eddington luminosity, where the efficiency for release of gravitational binding energy was taken to be that of a nonrotating black hole. Equation (5) reflects the fact that the ultimate energy source is the gravitational binding energy, which is unlocked by the interaction of the turbulent viscosity with the Keplerian shear. In a stationary disk

$$F = Q, \qquad (7)$$

with $F$ denoting the energy flux emerging from each side of the disk surface. Since in this work we do not solve for a detailed vertical structure, the flux $F$ is taken to be a vertical average. Since also $Q$, in equation (3), is represented in terms of vertically averaged quantities, we interpret now equation (7) to represent a relation between vertical averages. This is true up to factors of order unity that depend on the details of the vertical structure.

The vertically averaged equation of hydrostatic equilibrium (up to a factor of order unity which depends on the detailed vertical structure) is

$$P = \rho \Omega^2 h^2, \qquad (8)$$

where $P$ is the total pressure at the disk midplane. As seen below, turbulent convection is subsonic so the turbulent pressure can be neglected compared to the thermal pressure, so that

$$P = P_g + P_r, \qquad (9)$$

with the gas pressure, $P_g$

$$P_g = \frac{k_B}{m} \rho T \qquad (10)$$

and the radiation pressure, $P_r$

$$P_r = \frac{1}{3} a T^4, \qquad (11)$$

where $k_B$ is the Boltzman constant, m is the atomic mass per particle ($m = 0.59 m_p$ for solar abundances), $a$ is the blackbody constant, and $T$ is the midplane temperature. For convenience we introduce the parameter $\beta$ — the ratio between the gas pressure and the total pressure

$$\beta = \frac{P_g}{P}. \qquad (12)$$

## 3. Turbulent Convective Viscosity

The model for turbulence employed (Canuto *et al.* 1987) yields the turbulence spectrum in terms of $n(k)$ — the rate controlling the energy input, at wavenumber $k$, from the generating source into the turbulence. The turbulent viscosity, which is contributed by all the eddies comprising the turbulence spectrum, can nevertheless be expressed simply as

$$\nu_t = n(k_0) k_0^{-2}, \tag{13}$$

where $k_0$ is the smallest wavenumber (largest eddy) present in the turbulence spectrum. In principle, $n(k)$ depends also on the turbulence spectrum itself. However, Canuto *et al.* (1987) demonstrated that in the case of turbulent convection, it can be approximated by the growth rate of the unstable modes of the linearized equations. These authors found that the resulting turbulence spectrum and convective flux are in good agreement with those derived from a self-consistent renormalized $n(k)$, and with laboratory experiments on convection at high Rayleigh numbers.

Turbulent convection in accretion disks is likely to be affected by the interaction of the strong Keplerian shear with the eddies. In the turbulence model, employed here, the effects of the shear are introduced through the growth rate $n(k)$ which corresponds to convection in a sheared (differentially rotating) Keplerian disk (Goldreich & Schubert 1967; Canuto, Goldman, & Hubickyj 1984). The growth rate for the largest eddies (smallest wavenumbers) which determines the turbulent viscosity in equation (13), can be expressed in terms of a dimensionless growth rate $N$ multiplied by the inverse of the buoyancy timescale

$$n(k_0) = N \left( g_z \bar{\alpha} \delta \frac{T}{z} \right)^{1/2} = N \Omega \bar{\delta}^{1/2}. \tag{14}$$

where $N$ is the positive real part of the solution of a third order algebraic equation (Canuto *et al.* 1984). Its dependence on the physical parameters will be considered in §5. In equation (14), $g_z = \Omega^2 z$ is the vertical gravitational acceleration at height $z$ due to the central object, $\bar{\alpha}$ is the coefficient of thermal volume expansion at constant pressure (Chandrasekhar 1967)

$$\bar{\alpha} = \frac{1}{V}\left(\frac{\partial V}{\partial T}\right)_p = T^{-1}\frac{(4-3\beta)}{\beta}. \qquad (15)$$

The dimensionless superadiabatic temperature gradient $\delta$ is given by

$$\delta = \left\langle -\frac{z}{T}\left\{\left(\frac{dT}{dz}\right) - \left(\frac{dT}{dz}\right)_{ad}\right\}\right\rangle, \qquad (16)$$

with the angular brackets denoting vertical average. The parameter $\bar{\delta}$ appearing in the right handside of equation (14) is given by

$$\bar{\delta} = \frac{(4-3\beta)}{\beta}\delta. \qquad (17)$$

For gas pressure dominated regions $\bar{\delta} = \delta$ while for radiation pressure dominated regions $\bar{\delta} = 4\beta^{-1}\delta \gg \delta$. All the quantities in equation (14) represent vertical averages. The wavenumber $k_0$ is expressible in terms of the disk half-width $h$

$$k_0 = h^{-1}\pi(1+x)^{1/2}, \qquad (18)$$

where $x = k_p^2/k_z^2$ is a measure of the anisotropy of the largest eddies, $k_p$ and $k_z$ are the horizontal and vertical wavenumbers respectively, so that $k^2 = k_p^2 + k_z^2$ with $k_z h = \pi$.

Using equations (14) and (18) in equation (13) yields

$$\nu_t = \frac{N}{\pi^2(1+x)}\bar{\delta}^{1/2}\Omega h^2. \qquad (19)$$

Before discussing the dependence of $N$ on the physical parameters, we present the expression for the convective flux.

## 4. Turbulent Convective Flux

Since in the present case convection is the generating mechanism of the disk turbulence, it is conceivable that the convective flux could be important in the transfer of the released energy to the disk surface. Therefore, in equation (7) the vertically averaged flux is taken to be the sum of the radiative and convective fluxes

$$F = F_r + F_c . \qquad (20)$$

The radiative flux $F_r$ is given by

$$F_r = \frac{4acT^4}{3\tau}, \qquad (21)$$

with $\tau = \kappa\rho h$ the optical depth ($T$, $\rho$, and the opacity $\kappa$ are midplane values). The vertically averaged convective flux, $F_c$ is given by (Canuto et al. 1987)

$$F_c = c_p\rho \frac{1}{g_z\bar{\alpha}} A\pi^2 n^3(k_0) k_0^{-2}, \qquad (22)$$

with $A$ a dimensionless constant that depends on the strength of convection, as detailed in § 5. The specific heat at constant pressure $c_p$ is (Chandrasekhar 1967)

$$c_p = \frac{\Gamma_2}{\Gamma_2 - 1} \frac{k_B}{m} \frac{(4 - 3\beta)}{\beta^2} = \frac{\Gamma_2}{\Gamma_2 - 1} \bar{\alpha} \frac{P}{\rho} = \frac{\Gamma_2}{\Gamma_2 - 1} \bar{\alpha} \Omega^2 h^2, \qquad (23)$$

where $\Gamma_2$, denotes the second adiabatic index — equation (8) has been used to obtain the third equality. From equations (14), (15), (18), (22), and (23) follows that

$$F_c = \frac{A}{1+x} \frac{\Gamma_2}{\Gamma_2 - 1} \rho h^3 \Omega^3 N^3 \bar{\delta}^{3/2}. \qquad (24)$$

The fraction of the convective flux out of the total flux, $\xi_0$, can be obtained from equations (3), (19), and (24):

$$\xi_0 = \frac{F_c}{Q} = \frac{4}{9} A\pi^2 N^2 \frac{\Gamma_2}{\Gamma_2 - 1} \bar{\delta}. \qquad (25)$$

Before applying equations (19) and (24) to the disk equations, we consider below the dependence of the dimensionless growth rate, $N$, on the physical parameters of the disk.

## 5. The Limits of Strong and Moderate Convection

Using the standard expressions for the kinematic radiative viscosity (Weinberg 1971) and for the kinematic plasma viscosity (Spitzer 1962) we find that, for the physical conditions typical to accretion disks, both are much smaller than the radiative conductivity. Thus, the Prandtl number (the dimensionless ratio between the kinematic viscosity and the radiative conductivity) is very small compared to unity. Only for an extremely high radiation pressure is the radiative viscosity comparable to the radiative conductivity. In the limit of very small Prandtl number, the dimensionless growth rate $N$ depends only on the value of $S$— the dimensionless product of the Rayleigh and the Prandtl numbers (Canuto et al. 1984). For a thin disk

$$S = g_z \bar{\alpha} \delta \frac{T}{z} h^4 \chi^{-2} = \Omega^2 h^4 \bar{\delta} \chi^{-2}, \tag{26}$$

where $\chi$ is the radiative conductivity

$$\chi = \frac{F_r h}{c_p \rho T}. \tag{27}$$

The possible solutions of the equation for $N(S)$ (Canuto et al. 1984) can be classified according to the strength of the convection, measured by the dimensionless parameter $S^*$,

$$S^* = \frac{S^{1/2} N(S)}{\pi^2 (1+x)}. \tag{28}$$

Strong convection is obtained in the limit $S^* \gg 1$, weak convection corresponds to $S^* \ll 1$ and moderate convection to $S^* \sim 1$. Using equations (19) and (26) in equation (28) one finds

$$S^* = \frac{\nu_t}{\chi}, \tag{29}$$

which has a simple physical interpretation as the ratio between the timescales characterizing the radiative transport and the turbulence. Using now equation (3) to express $\nu_t$ in terms of $Q$ and equation (27) to express $\chi$ in terms of $F_r$, as well as equations (15) and (23), one finds

$$S^* = \frac{4}{9} \frac{\Gamma_2}{\Gamma_2 - 1} \frac{(4 - 3\beta)}{\beta} \frac{Q}{F_r} = \frac{4}{9} \frac{\Gamma_2}{\Gamma_2 - 1} \frac{(4 - 3\beta)}{\beta} (1 - \xi_0)^{-1}, \tag{30}$$

where $\xi_0$ is the ratio of convective to total flux, see equation (25). Equation (30) must be satisfied in any self-consistent disk solution in which the turbulent viscosity is due to convection. In radiation pressure dominated regions ($\beta \ll 1$) strong convection is guaranteed, even if the convective flux were small compared to the radiative flux (small $\xi_0$). In gas pressure dominated regions($\beta \to 1$) equation (30) yields $S^* = 1.1(1-\xi_0)^{-1}$, for $\Gamma_2 = 5/3$. Thus, in this case convection is either moderate or strong depending on the value of $\xi_0$. Strong convection occurs only for $\xi_0$ close to unity, namely only when the convective flux is much larger than the radiative flux. In each of these two limits we would express $N$ and $x$ in terms of $\delta$, so that both the turbulent viscosity and the convective flux will depend only on $\delta$. Practically, it will be more convenient to use equation (25) to express $\delta$ in terms of $\xi_0$, and use the latter as the parameter in the various expressions.

### 5.1. Strong Convection

In the strong convection limit one has (Canuto *et al.* 1984)

$$N = \left(\frac{x - \bar{\delta}^{-1}}{1 + x}\right)^{1/2} \tag{31}$$

For a given superadiabatic gradient there are many possible unstable modes, corresponding to different values of the anisotropy parameter $x$, each resulting in a different value for the turbulent viscosity. Generally, the latter will be a superposition over the different $x$ values. Since the disk structure will be dominated by those modes that contribute most to the turbulent viscosity and hence to the energy production, we approximate the superposition by that $x$ which maximizes the turbulent viscosity for a given superadiabatic gradient. From equations (19) and (31) then follows that

$$x = \frac{1}{2}(1 + 3\bar{\delta}^{-1}) \tag{32}$$

for which equation (31) yields

$$N = 3^{-1/2} \tag{33}$$

so that the turbulent viscosity, equation (19), becomes

$$\nu_t = 0.039 \frac{\bar{\delta}^{3/2}}{1 + \bar{\delta}} \Omega h^2. \tag{34}$$

In the strong-convection limit we find, using equation (22) with equation (52) of Canuto *et al.* (1987), that the parameter $A$ appearing in the expression for the convective flux, equation (24) is $\sim 3$. For the above values of $N$ and $A$, equation (25) yields

$$\xi_0 = 4.4 \frac{\Gamma_2}{\Gamma_2 - 1} \bar{\delta}, \tag{35}$$

which for radiation pressure dominated regions ($\Gamma = 4/3$) is

$$\xi_0 = 17.6\bar{\delta}, \tag{36}$$

while for gas pressure dominated regions ($\Gamma = 5/3$),

$$\xi_0 = 11.0\bar{\delta}. \tag{37}$$

Since $\xi_0 \leq 1$ by definition, it follows that for radiation pressure $\bar{\delta} \leq 0.057$ and for gas pressure $\bar{\delta} \leq 0.09$. Thus, the temperature gradient is close to adiabatic; in particular so in the case of radiation pressure for which $\delta \ll \bar{\delta}$. Such a situation is indeed expected for strong convection which acts to reduce the temperature gradient. The corresponding values of $x$ are quite large: $x \gtrsim 25$ for radiation pressure and $x \gtrsim 17$ for gas pressure dominated regions. This implies that the largest turbulent eddies are anisotropic with the horizontal dimension smaller than the vertical dimension by a factor of $x^{1/2}$ ($\gtrsim 4-5$). Such anisotropy, resulting from the Coriolis force, is a general feature of any three dimensional turbulence in a differentially rotating disk (Goldman 1991).

Equation (34) can be used to define an effective $\alpha$ parameter

$$\alpha_{eff} \equiv \frac{\nu_t}{\Omega h^2} = 0.04 \frac{\bar{\delta}^{3/2}}{1+\bar{\delta}}. \tag{38}$$

From equations (36) and (37) follows that $\alpha_{eff} \lesssim 5 \times 10^{-4}$ for radiation pressure dominated regions and $\lesssim 10^{-3}$ for gas pressure dominated regions. The small values of $\alpha_{eff}$ result from the small value of the turbulent velocity compared to the sound velocity (see eq. [44])) as well as from the large anisotropy of the eddies with the horizontal dimension much smaller than the disk half-width.

### 5.2. Moderate Convection

As noted above, in radiation pressure regions convection is always strong but in gas pressure regions it can be also moderate. From Canuto *et al.* (1984) we find a solution for $N$ for which $S^* \sim 1$

$$N \sim \left(\frac{1}{6}\frac{x - \frac{2}{3}\bar{\delta}^{-1}}{1+x}\right)^{1/2}. \tag{39}$$

Therefore, the maximal turbulent viscosity is obtained for an anisotropy parameter

$$x \sim \frac{1}{2}(1 + 2\bar{\delta}^{-1}) \tag{40}$$

yielding

$$N \sim (18)^{-1/2} \sim 0.24. \tag{41}$$

Using equations (52) and (56) of Canuto *et al.* (1987), we find that for moderate convection, the dimensionless parameter $A \sim 1$. For the above values of $A$ and $N$ and for $\Gamma = 5/3$, equation (25) results in

$$\xi_0 \sim 0.6\bar{\delta}. \tag{42}$$

Use of equations (40), (41) and (42) in equation (19) yields the turbulent viscosity

$$\nu_t \sim 0.016 \frac{\bar{\delta}^{3/2}}{\frac{2}{3}+\bar{\delta}}\Omega h^2 = 0.05 \frac{\xi_0^{3/2}}{1+2.5\xi_0}\Omega h^2. \tag{43}$$

We note that moderate convection is obtained when $S^*$ is of order unity. Therefore, all the numerical values, in the relations above, should be regarded as representative, and could actually differ by factors of order unity.

Equation (42) implies that $\delta$ could be larger than in the case of strong convection (for gas pressure regions $\bar{\delta} \sim \delta$). For such larger values of $\delta$, the anisotropy parameter, given by equation (40), is significantly smaller than in the case of strong convection. The effect of the smaller anisotropy parameter overcomes the decrease in $N$ so that the effective $\alpha$- parameter can be somewhat larger than in the case of strong convection. For example, $\xi_0 = 0.3$ yields $\delta \sim 0.5$, implying $x \sim 2.5$ and $\alpha_{eff} \sim 5 \times 10^{-3}$.

## 6. Disk Structure — Strong Convection

The expressions for the turbulent viscosity and for the turbulent convective flux, in the strong-convection limit, are used below to solve for the disk structure. This is done for three different regions: radiation pressure dominated, gas pressure dominated with electron scattering opacity and gas pressure dominated, free-free absorption opacity, blackbody region. Given the mass of and the distance from the central object, and the accretion rate, we obtain the disk half-width $h$, the midplane values of the temperature $T$ and the density $\rho$, and the ratio of the convective to total flux, as functions of the dimensionless superadiabatic temperature gradient.

Before doing so we note that the turbulent pressure need not be included in the hydrostatic equilibrium equation since the turbulent velocity is much smaller than the thermal velocity. The turbulent velocity (rms value), obtained using Canuto et al. (1987), is

$$v_t \sim \left[A\pi^2 n^2(k_0)k_0^{-2}\right]1/2 = A^{1/2}\frac{N}{(1+x)^{1/2}}\bar{\delta}^{1/2}\Omega h, \tag{44}$$

implying that the ratio of the turbulent to the thermal velocity ($\sim \Omega h$) is $\lesssim 0.06$ for both radiation pressure and gas pressure dominated regions.

### 6.1. Radiation Pressure Region

In this region, $\beta \ll 1$, $\kappa = \kappa_{es} = 0.4\text{g}^{-1}\text{cm}^2$, and the strong convection limit applies regardless of the value of $\xi_0$. Expressing the total pressure as

$$P = \frac{1}{3}aT^4\frac{1}{1-\beta}, \tag{45}$$

the hydrostatic equilibrium equation, equation (8) takes the form

$$\frac{1}{3}aT^4\frac{1}{1-\beta} = \rho\Omega^2 h^2. \tag{46}$$

We use equation (25) to express the convective flux in terms of $\xi_0$ and $Q$, so that equation (20) becomes,

$$F_r = (1-\xi_0)Q, \tag{47}$$

which upon using equations (5) and (21) yields

$$\frac{4acT^4}{3\kappa_{es}\rho h} = (1-\xi_0)\frac{3}{8\pi}\Omega^2\dot{M}\phi(r) = 1.2 \times 10^{20}(1-\xi_0)L_*M_8^{-1}r^{-3}\,erg\,s^{-1}. \tag{48}$$

The angular momentum equation, equation (4) with $\nu_t$ given by equation (34) results in

$$\dot{M}\phi(r) = 0.753\rho\Omega h^3 \frac{\bar{\delta}^{3/2}}{1+\bar{\delta}} = 0.01\rho\Omega h^3 \frac{\xi_0^{3/2}}{1+0.057\xi_0}, \quad (49)$$

where the second equality follows from equation (36). In what follows the denominator in the right hand side of equation (49) will be approximated by 1 (since $\xi_0 \leq 1$).

Equations (46), (48) and (49), with equation (6), can be solved to obtain $T$, $\rho$ and $h$ as functions of $\xi_0$ and the distance from the central object. Equation (36) can be used to express $\xi_0$ in terms of $\bar{\delta}$. Combining equations (46) and (48) yields

$$h = (1 \times 10^{14} cm) \, M_8 L_* (1-\beta)^{-1} (1-\xi_0), \quad (50)$$

so that

$$\frac{h}{R} = 6.63 L_* (1-\beta)^{-1} (1-\xi_0) r^{-1}. \quad (51)$$

Substituting $h$ from equation (50) and $\Omega$ from equation (2), in equation (49) yields the midplane density $\rho$

$$\rho = (1.2 \times 10^{-11} g cm^{-3}) \, M_8^{-1} L_*^{-2} (1-\beta)^3 (1-\xi_0)^{-3} \xi_0^{-3/2} r^{3/2}, \quad (52)$$

which when used with equation (50) yields the optical depth for electron scattering

$$\tau_{es} = \kappa_{es}\rho h = 500 L_*^{-1} (1-\beta)^2 (1-\xi_0)^{-2} \xi_0^{-3/2} r^{3/2}. \quad (53)$$

The midplane temperature $T$ is now obtained from equations (48) and (53)

$$T = (3.7 \times 10^6 K) \, M_8^{-1/4} (1-\beta)^{1/2} (1-\xi_0)^{-1/4} \xi_0^{-3/8} r^{-3/8}, \quad (54)$$

Equations (10), (11), (52), and (54) yield

$$\frac{P_r}{P_g} = 7.6 \times 10^7 (1-\xi_0)^{9/4} \xi_0^{3/8} L_*^2 M_8^{1/4} (1-\beta)^{-3/2} r^{-21/8}. \quad (55)$$

Therefore, a necessary condition for the radiation pressure region to exist is a value of $\xi_0$ not too close to unity or to zero. Thus, demanding $\beta = 0.2$ for $r = 20$ implies $2 \times 10^{-11} L_*^{-16/3} M_8^{-2/3} \lesssim \xi_0 \lesssim 1 - 0.016 L_*^{-8/9} M_8^{-1/9}$.

Formally, for $\xi_0 \ll 1$ the above expressions for $h$, $\rho$, $\tau_{es}$, and $T$, have the same form as in a $\alpha$-disk with a small value of $\alpha_{eff}$ (resulting from the small $\xi_0$). Unlike in the $\alpha$-disk model, the effective value of $\alpha$ is determined directly by the local physical conditions. However, as shown in § 8.5, in the radiation pressure region $\xi_0 \not\ll 1$. For $\xi_0 \not\ll 1$, the present convective disk is geometrically thinner and optically thicker the larger $\xi_0$ is. Note that for $\xi_0 \to 1$, equation (51) implies that the disk is geometrically-thin at small $r$ values even for super-Eddington accretion rates, $L_* \phi(r)^{-1} > 1$. An increase in $\xi_0$ causes also an increase of $T$ but the dependence is quite weak, see equation (54). In § 8.5 we present an estimate for the dependence of $\xi_0$ on the local disk parameters.

### 6.2. Gas Pressure Region with Electron Scattering Opacity

We consider first the limit of strong convection which in gas pressure dominated regions requires that $\xi_0 \gtrsim 0.9$, i.e. a convective flux which is at least an order of magnitude larger than the radiative flux. Now, $\kappa = \kappa_{es}$ but $1 - \beta \ll 1$ and it is convenient to express the total pressure as

$$P = \frac{k_B}{m\beta} \rho T, \tag{56}$$

which, when combined with equation (8), results in

$$\frac{k_B}{m\beta} T = \Omega^2 h^2. \tag{57}$$

Equation (48) is unchanged in the present case. Using equation (37), the second equality in equation (49) becomes

$$\dot{M}\phi(r) = 0.02 \rho \Omega h^3 \frac{\xi_0^{3/2}}{1 + 0.09 \xi_0} \tag{58}$$

and, here too, the denominator will be taken as equal 1. Combining equations (48), (57) and (58), with equation (6), yields the midplane temperature $T$,

$$T = (1.5 \times 10^8 K)\, M_8^{-1/5} L_*^{2/5} \beta^{1/5} (1 - \xi_0)^{1/5} \xi_0^{-3/10} r^{-9/10}. \tag{59}$$

Further combination of equations (57) and (59) yields the half width of the disk

$$h = (7.3 \times 10^{10} cm) \, M_8^{9/10} L_*^{1/5} \beta^{-2/5} (1 - \xi_0)^{1/10} \xi_0^{-3/20} r^{21/20}, \tag{60}$$

so that

$$\frac{h}{R} = 4.9 \times 10^{-3} M_8^{-1/10} L_*^{1/5} \beta^{-2/5} (1 - \xi_0)^{1/10} \xi_0^{-3/20} r^{1/20}. \tag{61}$$

The electron scattering optical depth is obtained from equations (48) and (59)

$$\tau_{es} = 1.3 \times 10^9 L_*^{3/5} M_8^{1/5} \beta^{4/5} (1 - \xi_0)^{-1/5} \xi_0^{-6/5} r^{-3/5}. \tag{62}$$

The midplane density $\rho$ is obtained directly from equations (60) and (62)

$$\rho = (4.5 \times 10^{-2} g cm^{-3}) \, L_*^{2/5} M_8^{-7/10} \beta^{6/5} (1 - \xi_0)^{-3/10} \xi_0^{-21/20} r^{-33/20}. \tag{63}$$

The disk is geometrically thinner, optically thicker and has a lower midplane temperature the larger is $\xi_0$.

### 6.3. Gas Pressure Blackbody Region

In this region the dominant opacity is due to free-free absorption

$$\kappa_{ff} = k_o \rho T^{-7/2}, \tag{64}$$

where

$$k_o = 6.45 \times 10^{22} C_{bf} g^{-2} cm^6 K^{7/2} \tag{65}$$

and $C_{bf}$ is the bound-free enhancement factor which equals $\sim 30$ for solar abundances.

As in the case of electron scattering opacity we consider first the strong-convection limit. Equations (56), (57) and (58) are unchanged in this case. The only change is in equation (48) where the opacity is now $\kappa_{ff}$ instead of $\kappa_{es}$. Equations (57) and (58) can be use to express $h$ and $\rho$ in terms of $T$, so that $\kappa_{ff}$ of equation (64) can also be expressed by $T$. Defining

$$\xi_{bb} = \left(\frac{C_{bf}}{30}\right)^{1/20} (1 - \xi_0)^{1/20} \xi_0^{-3/20}$$

equation (48), with $\kappa_{ff}$ instead of $\kappa_{es}$ now yields

$$T = (2.6 \times 10^7 K) \, L_*^{3/10} M_8^{-1/5} \beta^{1/4} \xi_{bb}^2 r^{-3/4}. \tag{66}$$

Substituting this $T$ into equation (57) yields

$$h = (3.0 \times 10^{10} cm) \, M_8^{9/10} L_*^{3/20} \beta^{-3/8} \xi_{bb} r^{9/8} \tag{67}$$

so that

$$\frac{h}{R} = 2.0 \times 10^{-3} M_8^{-1/10} L_*^{3/20} \beta^{-3/8} \xi_{bb} r^{1/8}. \tag{68}$$

Finally, equations (58) and (67) result in

$$\rho = (0.21 g cm^{-3}) \, M_8^{-7/10} L_*^{11/20} \beta^{9/8} \xi_{bb}^{-3} \xi_0^{-3/2} r^{-15/8} \tag{69}$$

and

$$\tau_{es} = 2.8 \times 10^9 M_8^{1/5} L_*^{7/10} \beta^{3/4} \xi_{bb}^{-2} \xi_0^{-3/2} r^{-3/4} \tag{70}$$

## 7. Disk Structure — Moderate Convection

As already noted, contrary to the radiation pressure region in which convection is always strong, in the gas pressure region convection can be either strong or moderate. The first possibility was addressed in § 6. In what follows we consider the second possible class of solutions for the gas pressure region — that of moderate convection. In this case too, equation (44) implies that the turbulent velocity is much smaller than the thermal velocity thus the turbulent pressure need not be taken into account.

Equations (48) and (57) are unchanged while $\nu_t$ is now given by equation (43) and $\xi_0$ is related to $\bar{\delta}$ by equation (42). Therefore, instead of equation (58) the angular momentum equation will be

$$\dot{M}\phi(r) \sim 0.45 \rho \Omega h^3 \frac{\bar{\delta}^{3/2}}{1 + 1.5\bar{\delta}} \sim \rho \Omega h^3 \frac{\xi_0^{3/2}}{1 + 2.5\xi_0}, \tag{71}$$

where equation (42) was used to obtain the second equality. We wish to stress again that since for moderate convection $S^* \sim 1$ up to a factor of order unity, the relation between $\xi_0$ and $\delta$ (eq. [42]) as well as equation (71) can also vary by factors of order unity.

### 7.1. Gas Pressure Electron Scattering Region

Repeating the same steps as in § 6.2., and defining

$$\xi_* = (1 - \xi_0)^{1/5} \xi_0^{-3/10} (1 + 2.5\xi_0)^{1/5} \tag{72}$$

we obtain now

$$T = (6.9 \times 10^7 K)\, M_8^{-1/5} L_*^{2/5} \beta^{1/5} \xi_* r^{-9/10}, \tag{73}$$

$$h = (4.9 \times 10^{10} cm)\, M_8^{9/10} L_*^{1/5} \beta^{-2/5} \xi_*^{1/2} r^{21/20}, \tag{74}$$

$$\frac{h}{R} = 3.3 \times 10^{-3} M_8^{-1/10} L_*^{1/5} \beta^{-2/5} \xi_*^{1/2} r^{1/20}, \tag{75}$$

$$\tau_{es} = 5.7 \times 10^7 L_*^{3/5} M_8^{1/5} \beta^{4/5} (1-\xi_0)^{-1} \xi_*^4 r^{-3/5}, \tag{76}$$

$$\rho = (2.9 \times 10^{-3} g cm^{-3})\, L_*^{2/5} M_8^{-7/10} \beta^{6/5} (1-\xi_0)^{-1} \xi_*^{7/2} r^{-33/20}. \tag{77}$$

### 7.2. Gas Pressure Blackbody Region

Repeating the same steps as in § 6.3. we obtain now

$$T = (1.2 \times 10^7 K)\, \beta^{1/4} \left(\frac{C_{bf}}{30}\right)^{1/10} L_*^{3/10} M_8^{-1/5} \xi_* r^{-3/4}, \tag{78}$$

$$h = (2.0 \times 10^{10} cm)\, M_8^{9/10} L_*^{3/20} \beta^{-3/8} \left(\frac{C_{bf}}{30}\right)^{1/20} \xi_*^{1/2} r^{9/8}, \tag{79}$$

$$\frac{h}{R} = 1.35 \times 10^{-3} M_8^{-1/10} L_*^{3/20} \beta^{-3/8} \left(\frac{C_{bf}}{30}\right)^{1/20} \xi_*^{1/2} r^{1/8}, \tag{80}$$

$$\rho = (1.4 \times 10^{-2} g cm^{-3}) M_8^{-7/10} L_*^{11/20} \beta^{9/8} \left(\frac{C_{bf}}{30}\right)^{-3/20} (1-\xi_0)^{-1/2} \xi_*^{7/2} r^{-15/8}, \tag{81}$$

$$\tau_{es} = 1.1 \times 10^8 M_8^{1/5} L_*^{7/10} \beta^{3/4} \left(\frac{C_{bf}}{30}\right)^{-1/10} (1-\xi_0)^{-1/2} \xi_*^4 r^{-3/4}. \tag{82}$$

## 8. Determination of $\xi_0$ in the various Regions

The solutions in the various regions of the disk depend on $\xi_0$— the fraction of the convective flux out of the total flux. This is determined in each region by $\delta$— the superadiabatic temperature gradient. Obviously, the latter two are known once the detailed vertical structure is known. This however requires (see §1) a considerably more complex turbulence model than the one considered here. Thus, we wish to obtain an approximate determination of $\delta$, and consequently of $\xi_0$ within the framework of the vertically averaged approach. First we approximate the vertical average of the dimensionless temperature gradient by

$$\left\langle -\frac{z}{T}\frac{dT}{dz} \right\rangle \sim \frac{h}{T}\frac{T-T_s}{h} = 1 - \frac{T_s}{T}; \tag{83}$$

with $T_s$ denoting the surface temperature at the disk top, $z = h$. Next, the vertically averaged dimensionless adiabatic temperature gradient can be approximated, making use of the hydrostatic equilibrium equation (8), as

$$\left\langle -\frac{z}{T}\frac{dT}{dz} \right\rangle_{ad} \sim \frac{\Gamma_2 - 1}{\Gamma_2} = 1 - \frac{1}{\Gamma_2} \tag{84}$$

Thus, from equation (16) it follows that

$$\delta \sim \frac{1}{\Gamma_2} - \frac{T_s}{T}, \tag{85}$$

or equivalently

$$\frac{T_s}{T} \sim \frac{1}{\Gamma_2} - \delta \tag{86}$$

We note that, having been derived by an approximate vertically averaging, equations (85) and (86) are probably correct only up to factors of order unity. Even so, we expect the results to be representative of those corresponding to a more complex z-dependent modeling of the turbulence.

Above the top of the convective disk, there is a surface layer (of width $\Delta h << h$) in which the flux is purely radiative. We further assume that this layer is isothermal at a temperature $T_s$. If this layer is optically thick to absorption, or absorption modified by electron scattering, then the flux emerging from the disk can be represented by a blackbody or modified blackbody with a temperature $T_s$. In this case equation (86), together with the disk equations, would enable the determination of all the disk variables, including $\delta$ (or equivalently $\xi_0$). In what follows, we implement the above procedure to the various regions of the disk.

### 8.1. Gas Pressure Blackbody Region — Strong Convection

In the strong convection limit $\delta$ and $\xi_0$ are related by equation (37). The latter relation together with equation (86) yields, for $\Gamma_2 = 5/3$

$$\frac{T_s}{T} \sim 0.6 - 0.09\xi_0 \tag{87}$$

In the blackbody region, $T_s$ is determined by assuming a blackbody spectrum for the flux emerging from the disk surface

$$T_s = \left(\frac{4Q}{ac}\right)^{1/4} = 1.2 \times 10^6 K \; L_*^{1/4} M_8^{-1/4} r^{-3/4}, \tag{88}$$

where equation (5) has been used to obtain the second equality. Dividing this $T_s$ by $T$ of equation (66), and substituting in equation (87) yields an equation for $\xi_0$ (which is equivalent to an equation for $\delta$)

$$(0.6 - 0.09\xi_0)(1 - \xi_0)^{1/10} \xi_0^{-3/10} = 4.6 \times 10^{-2} M_8^{-1/20} L_*^{-1/20} \beta^{-1/4} \left(\frac{C_{bf}}{30}\right)^{-1/10}. \tag{89}$$

resulting in

$$1 - \xi_0 \sim 3.8 \times 10^{-11} M_8^{-1/2} L_*^{-1/2} \beta^{-5/2} \left(\frac{C_{bf}}{30}\right)^{-1}, \tag{90}$$

Thus, the convective flux is much larger than the radiative flux. With this value of $\xi_0$, we obtain from equation (37) that $\delta = 0.09$, and equation (87) yields $T_s/T = 0.51$. Substituting the above value of $\xi_0$ in equations (66)–(70) results in

$$T = (2.4 \times 10^6 K) \, L_*^{1/4} M_8^{-1/4} r^{-3/4}, \tag{91}$$

$$\frac{h}{R} = 6.1 \times 10^{-4} M_8^{-1/8} L_*^{1/8} \beta^{-1/2} r^{1/8}, \tag{92}$$

$$\tau_{es} = 3 \times 10^{10} M_8^{1/4} L_*^{3/4} \beta r^{-3/4}, \tag{93}$$

$$\rho = (8.1 gcm^{-3}) \, M_8^{-5/8} L_*^{5/8} \beta^{3/2} r^{-15/8}. \tag{94}$$

This disk solution would have been obtained had we started with pure convective transfer ($\xi_0 = 1$) and employed equations (56), (57), (58), (87), and (88).

The line marked BBS in figure 1 shows the $L - \tau_{es}$ relation for the strong convection limit in the gas pressure, blackbody region. The positive slope implies that this solution is secularily stable.

In order to find the boundary between the blackbody and electron scattering regions we evaluate the ratio between the free-free and electron scattering opacities. From equations (64), (65), (91), and (94) follows that

$$\frac{\kappa_{ff}}{\kappa_{es}} = 2.1 \times 10^3 \left(\frac{C_{bf}}{30}\right) M_8^{1/4} L_*^{-1/4} \beta^{3/2} r^{3/4}, \tag{95}$$

which is larger than unity for all values of $r$. Therefore, it is possible for this solution to encompass the entire gas pressure region. As will be shown in § 8.2, there is indeed no gas pressure dominated electron scattering region, in the strong convection limit.

In $\alpha$-disks there is a radiation pressure dominated region for small values of $r$, when $L_* \sim 1$. Could the entire disk, in the present case, be a gas pressure dominated blackbody? To find out we compute the ratio between the gas and radiation pressures, in the gas pressure blackbody region. From equations (10), (11), (91) and (94) we obtain

$$\frac{P_g}{P_r} = 3.25 \times 10^4 L_*^{-1/8} M_8^{1/8} \beta^{3/2} r^{3/8}. \tag{96}$$

Therefore, the disk could be gas pressure dominated blackbody for all values of r, even when $L_* \sim 1$. Note that unlike in the $\alpha$-disk, an increase of the mass of the central object favors gas pressure dominance.

## 8.2. Gas Pressure Electron Scattering Region — Strong Convection

In the electron scattering region, we apply the modified blackbody approximation to the emerging surface flux (Rybicki & Lightman 1979)

$$Q = \frac{ac}{4} T_s^4 \left(\frac{\kappa_{ff}}{\kappa_{es}}\right)^{1/2}, \tag{97}$$

which is valid when $\kappa_{es} \gg \kappa_{ff}$. The absorption opacity $\kappa_{ff}$ is evaluated by substituting in equation (64) the surface temperature and the midplane density. One obtains from equations (5), (63), (64) and (97) the value of $T_s$ which when divided by $T$ of equation (59), yields

$$\frac{T_s}{T} = 2.8 \times 10^{-3} \beta^{-7/15} \left(\frac{C_{bf}}{30}\right)^{-2/9} L_*^{-2/45} M_8^{-4/45} (1-\xi_0)^{-2/15} \xi_0^{8/15} r^{-1/15}. \tag{98}$$

Using this in equation (87) yields an equation similar to equation (89)

$$(0.6 - 0.09\xi_0)(1-\xi_0)^{2/15} \xi_0^{-8/15} = 2.8 \times 10^{-3} \beta^{-7/15} \left(\frac{C_{bf}}{30}\right)^{-2/9} L_*^{-2/45} M_8^{-4/45} r^{-1/15}. \tag{99}$$

Equation (99) yields a value of $\xi_0$ which is extremely close to 1

$$1 - \xi_0 \sim 1 \times 10^{-17} M_8^{-2/3} L_*^{-1/3} \beta^{-7/2} \left(\frac{C_{bf}}{30}\right)^{-5/3} r^{-1/2}. \tag{100}$$

The ratio $\kappa_{ff}/\kappa_{es}$, is obtained from equations (64) and (65), with $T$ and $\rho$ given by equations (59) and (63), respectively, and $\xi_0$ given by equation (100)

$$\frac{\kappa_{ff}}{\kappa_{es}} = 4.6 \times 10^{11} \left(\frac{C_{bf}}{30}\right)^{8/3} M_8^{2/3} L_*^{-2/3} \beta^4 r^2, \tag{101}$$

The above ratio is larger than unity for all values of $r$. This result is inconsistent with the definition of the electron scattering region. Since we are dealing with a thin disk, it is of interest to repeat the above procedure with $\kappa_{ff}$ in equation (97) evaluated in terms of the mid-plan rather than the surface temperature. Doing so yields again $\xi_0$ extremely close to unity, and the conclusion remains unchanged. Thus, we conclude that in the strong-convection limit there is no gas pressure dominated electron-scattering region and the entire gas pressure region is a blackbody, described by equations (91)-(94).

### 8.3. Gas Pressure Blackbody Region — Moderate Convection

We follow here the same procedure as in the case of strong convection, § 8.1. Equation (86) for $\Gamma_2 = 5/3$ yields, upon using equation (42) to express $\delta$ in terms of $\xi_0$,

$$\frac{T_s}{T} \sim 0.6 - 1.67\xi_0 \qquad (102)$$

From equations (72), (78), and (88) we obtain the ratio of $T_s/T$, which when substituted in equation (102) results in

$$(0.6 - 1.67\xi_0)(1-\xi_0)^{1/10}\xi_0^{-3/10}(1+2.5\xi_0)^{1/5} = 0.1 M_8^{-1/20} L_*^{-1/20} \beta^{-1/4} \left(\frac{C_{bf}}{30}\right)^{-1/10}. \qquad (103)$$

For given values of the various dimensionless parameters, equation (103) can be solved for $\xi_0$. Substitution of this $\xi_0$ into equations (78)-(82) will yield the values of the disk physical variables. However, it is not possible to obtain an analytical expression for $\xi_0$ in terms of the dimensionless parameters, as was done in § 8.1. There, it was possible because $\xi_0$ was very close to unity, for all relevant values of the dimensionless parameters.

The right hand side of equation (103) is insensitive to the values of the dimensionless parameters. For a representative case where they are set equal to unity the solution of equation (103) is

$$\xi_0 \sim 0.32 \qquad (104)$$

corresponding to $\delta \sim 0.53 \lesssim 1/\Gamma_2$. Substituting this $\xi_0$ in equations (78) – (82) results in

$$T = (1.8 \times 10^7 K)\, \beta^{1/4} \left(\frac{C_{bf}}{30}\right)^{1/10} L_*^{3/10} M_8^{-1/5} r^{-3/4}, \qquad (105)$$

$$h = (2.5 \times 10^{10} cm)\, M_8^{9/10} L_*^{3/20} \beta^{-3/8} \left(\frac{C_{bf}}{30}\right)^{1/20} r^{9/8}, \qquad (106)$$

$$\frac{h}{R} = 1.7 \times 10^{-3} M_8^{-1/10} L_*^{3/20} \beta^{-3/8} \left(\frac{C_{bf}}{30}\right)^{1/20} r^{1/8}, \qquad (107)$$

$$\rho = (7.4 \times 10^{-2} g cm^{-3})\, M_8^{-7/10} L_*^{11/20} \beta^{9/8} \left(\frac{C_{bf}}{30}\right)^{-3/20} r^{-15/8}, \qquad (108)$$

$$\tau_{es} = 7.4 \times 10^8 M_8^{1/5} L_*^{7/10} \beta^{3/4} \left(\frac{C_{bf}}{30}\right)^{-1/10} r^{-3/4}. \qquad (109)$$

The line marked BBM in figure 1 shows the $L - \tau_{es}$ relation for the moderate convection limit in the gas pressure, blackbody region. The positive slope of the curve implies that this solution is secularily stable.

To find the boundary between the electron scattering and blackbody regions, as determined in the blackbody region, we apply $T$ and $\rho$ of equations (105) and (108) to equation (64) and obtain

$$\frac{\kappa_{ff}}{\kappa_{es}} = 1.4 \times 10^{-2} \left(\frac{C_{bf}}{30}\right)^{1/2} L_*^{-1/2} \beta^{1/4} r^{3/4} \qquad (110)$$

implying that the blackbody region ($\kappa_{ff} > \kappa_{es}$) exists for

$$r > 280 \left(\frac{C_{bf}}{30}\right)^{-2/3} L_*^{2/3} \beta^{-1/3}, \qquad (111)$$

but see equation (120).

### 8.4. Gas Pressure Electron Scattering Region — Moderate Convection

Similarly to the case of strong convection, the surface temperature obtained from the modified blackbody approximation, equation (97), and the temperature given by equation (73) are substituted in equation (102), resulting in

$$(0.6 - 1.67\xi_0)(1 - \xi_0)^{2/15}\xi_0^{-8/15}(1 + 2.5\xi_0)^{16/45} =$$
$$1.1 \times 10^{-2}\beta^{-7/15}\left(\frac{C_{bf}}{30}\right)^{-2/9} L_*^{-2/45} M_8^{-4/45} r^{-1/15}. \tag{112}$$

Once the values of the various dimensionless parameters are specified, equation (112) can be solved for $\xi_0$. Substitution of this $\xi_0$ into equations (73)-(77) will yield the values of the disk physical variables. However, it is not possible to obtain an analytical expression for $\xi_0$ in terms of the dimensionless parameters, as was done in §8.2. There, it was possible because $\xi_0$ was very close to unity, for all relevant values of the dimensionless parameters.

Setting all the dimensionless parameters equal to unity, equation (112) yields

$$\xi_0 \sim 0.357 \tag{113}$$

corresponding to $\delta \sim 0.59 \lesssim 1/\Gamma_2$. We note that an increase of the right hand side of equation (112) by a factor of 10 yields $\xi_0 \sim 0.33$, while decreasing the right hand side results in $\xi_0$ larger than that of equation (113) but smaller than 0.36. For $\xi_0 = 0.357$ equations (73) – (77) become

$$T = (9.8 \times 10^7 K)\, M_8^{-1/5} L_*^{2/5} \beta^{1/5} r^{-9/10}, \tag{114}$$

$$h = (5.8 \times 10^{10} cm)\, M_8^{9/10} L_*^{1/5} \beta^{-2/5} r^{21/20}, \tag{115}$$

$$\frac{h}{R} = 3.9 \times 10^{-3} M_8^{-1/10} L_*^{1/5} \beta^{-2/5} r^{1/20}, \tag{116}$$

$$\tau_{es} = 3.6 \times 10^8 L_*^{3/5} M_8^{1/5} \beta^{4/5} r^{-3/5}, \tag{117}$$

$$\rho = (1.6 \times 10^{-2} g cm^{-3})\, L_*^{2/5} M_8^{-7/10} \beta^{6/5} r^{-33/20}. \tag{118}$$

The line marked ESM in figure 1 shows the $L - \tau_{es}$ relation for the moderate convection limit in the gas pressure, electron scattering region. The positive slope implies that this solution is secularily stable.

To find the boundary between the electron scattering and blackbody regions, we substitute $T$ and $\rho$ from equations (114) and (118) to equation (64) and obtain

$$\frac{\kappa_{ff}}{\kappa_{es}} = 8.3 \times 10^{-6} \left(\frac{C_{bf}}{30}\right) L_*^{-1} \beta^{1/2} r^{3/2} ; \tag{119}$$

implying that the electron scattering region ($\kappa_{ff} < \kappa_{es}$) exists for

$$r < 2.4 \times 10^3 \left(\frac{C_{bf}}{30}\right)^{-2/3} L_*^{2/3} \beta^{-1/3}. \tag{120}$$

In determining the boundary between the electron scattering and the blackbody gas pressure regions both equations (111) and (120) should be considered.

In order to find the boundary between the electron scattering gas pressure region and the radiation pressure region we evaluate the ratio between the gas and radiation pressures in the gas pressure electron scattering region. Employing equations (10), (11), (114) and (118) results in

$$\frac{P_g}{P_r} = 9.4 \times 10^{-4} L_*^{-4/5} M_8^{-1/10} \beta^{3/5} r^{21/20}. \tag{121}$$

Thus, $P_g > P_r$ for

$$r > 763 L_*^{16/21} M_8^{2/21} \beta^{4/7}. \tag{122}$$

### 8.5. Radiation Pressure Region

In the radiation pressure dominated region equation (36) implies that $\delta \ll \bar{\delta} < 0.057$, meaning that the disk is effectively adiabatic so that $\rho \propto T^3$. The density at the disk surface, $\rho_s$ is expressible as

$$\rho_s = \rho(T_s/T)^3 \tag{123}$$

In order for the modified blackbody approximation to apply, it is necessary that the effective optical depth for absorption of the surface layer

$$\tau_{eff,s} = \kappa_{es} \rho_s \Delta h \left(\frac{\kappa_{ff}}{\kappa_{es}}\right)^{1/2} \tag{124}$$

be larger than unity. Here $\Delta h$ ($\ll h$) is the width of the surface layer, and $\kappa_{ff}$ is evaluated in terms of $\rho_s$ and $T_s$.

In the surface layer the flux is purely radiative and constant. This means that the gradient of the radiative pressure divided by the density is constant. On the other hand the gravitational acceleration is proportional to $z$, thus only the (much smaller) gas pressure gradient divided by density is varying. Solving the set of coupled equations for the temperature and density we find that indeed the surface layer in this case is nearly isothermal, the density decreases as a gaussian, and $\Delta h \sim h(\beta T_s/T)^{1/2}$. Using this $\Delta h$ and equations (52), (53), (54), (64), and (123), equation (124) yields

$$\tau_{eff,s} = 1.1 L_*^{-8/15} M_8^{-1/60} (1-\beta)^{7/10} \left(\frac{C_{bf}}{30}\right)^{2/15} (1-\xi_0)^{-31/20} \xi_0^{-17/40} r^{31/40} \left(\beta \frac{T_s}{T}\right)^{1/2}. \tag{125}$$

Assuming for the moment that indeed this $\tau_{eff,s} > 1$, we employ equation (97) with equations (52), (54), (64), and (123) to obtain

$$\frac{T_s}{T} = 5.14 (1-\beta)^{-7/10} \left(\frac{C_{bf}}{30}\right)^{-2/15} L_*^{8/15} M_8^{1/60} (1-\xi_0)^{11/20} \xi_0^{17/40} r^{-31/40}. \tag{126}$$

from which

$$(1-\xi_0)\xi_0^{17/22} = 0.05 M_8^{-1/33} L_*^{-32/33} (1-\beta)^{14/11} \left(\frac{C_{bf}}{30}\right)^{8/33} r^{31/22} \left(\frac{T_s}{T}\right)^{20/11} \tag{127}$$

At each given radial distance, $r$, this constitutes an equation for $\xi_0$ as function of the accretion rate (expressed via $L_*$), the mass of the compact object and the local physical conditions. The latter are manifested through $\beta$, $C_{bf}$ and most importantly the ratio $T_s/T$. The estimate of equation (86), for $\Gamma_2 = 4/3$, yields $T_s/T \sim 0.75$. As already noted, this is only an order of magnitude estimate. For this value, and taking $r \sim 20$, there is a solution only for $L_* \gtrsim 1$. To clarify the situation further, let us use equation (127) and obtain the dependence of $\tau_{eff,s}$ on the value of $T_s/T$. Substituting $L_*$ from equation (127) into equation (125) yields

$$\tau_{eff,s} = 5.64\beta^{1/2} \left(\frac{T_s}{T}\right)^{-1/2} (1-\xi_0)^{-1} \tag{128}$$

As $\beta \ll 1$, for $T_s/T = .75$ the optical depth exceeds unity only if $\xi_0$ is close to one, which in turn by equation (127) requires $L_* \gtrsim 1$. Alternatively, substituting $(1-\xi_0)$ from equation (127) into equation (125) results in

$$\tau_{eff,s} = 1.6\beta^{1/2} \left(\frac{T_s}{T}\right)^{-51/22} L_*^{32/33} M_8^{1/33} (1-\beta)^{-14/11} \left(\frac{C_{bf}}{30}\right)^{-8/33} \xi_0^{17/22} \left(\frac{r}{20}\right)^{-31/22}. \tag{129}$$

leading to the same conclusion that $L_* \gtrsim 1$ is required. Moreover, substituting $(1-\xi_0)$ from equation (127) into equation (51) yields

$$\frac{h}{R} = 0.34 L_*^{1/33} (1-\beta)^{3/11} M_8^{-1/33} \xi_0^{-17/22} r^{9/22} \left(\frac{C_{bf}}{30}\right)^{8/33} \left(\frac{T_s}{T}\right)^{20/11} \tag{130}$$

which for $r = 20$ and $T_s/T = 0.75$ is $\sim 0.7$, so the disk is no longer geometrically thin. We conclude that in order for the modified blackbody approximation to be selfconsistent in this region, $T_s/T$ must be smaller. In the absence of a detailed vertical structure, the actual value is not known. To illustrate the effect of a smaller $T_s/T$ we adopt a value smaller by a factor of 10: 0.075. In this case equation (127) becomes

$$(1-\xi_0)\xi_0^{17/22} = 0.03 M_8^{-1/33} L_*^{-32/33} (1-\beta)^{14/11} \left(\frac{C_{bf}}{30}\right)^{8/33} \left(\frac{r}{20}\right)^{31/22} \tag{131}$$

and equations (129) and (130) yield

$$\tau_{eff,s} = 660\beta^{1/2} L_*^{32/33} M_8^{1/33} (1-\beta)^{-14/11} \left(\frac{C_{bf}}{30}\right)^{-8/33} \xi_0^{17/22} \left(\frac{r}{20}\right)^{-31/22}. \tag{132}$$

$$\frac{h}{R} = 0.01 L_*^{1/33} (1-\beta)^{3/11} M_8^{-1/33} \xi_0^{-17/22} \left(\frac{r}{20}\right)^{9/22} \left(\frac{C_{bf}}{30}\right)^{8/33} \tag{133}$$

For $r = 20$ equation (131) has a solution if $L_* \gtrsim 0.1$. The value of $\xi_0$ ranges from $\sim 0.45$ up to $\sim 0.97$ when $L_*$ varies from $\sim 0.1$ to $\sim 1$, respectively. For all these solutions the surface layer is optically thick for absorption (modified by electron scattering) and the disk is geometrically thin, even for $L_* \gtrsim 1$.

Substituting $(1 - \xi_0)$ from equation (131) into equations (52)-(55) results in

$$\rho = (3.5 \times 10^{-5} g cm^{-3}) \, M_8^{-10/11} L_*^{10/11} (1-\beta)^{-9/11} \xi_0^{9/11} \left(\frac{r}{20}\right)^{-30/11} \left(\frac{C_{bf}}{30}\right)^{-8/11}, \quad (134)$$

$$\tau_{es} = 4.6 \times 10^7 M_8^{2/33} L_*^{31/33} (1-\beta)^{-6/11} \xi_0^{1/22} \left(\frac{r}{20}\right)^{-29/22} \left(\frac{C_{bf}}{30}\right)^{-16/33}, \quad (135)$$

$$T = (2.9 \times 10^6 K) \, M_8^{-8/33} L_*^{8/33} (1-\beta)^{2/11} \left(\frac{C_{bf}}{30}\right)^{-2/33} \xi_0^{-2/11} \left(\frac{r}{20}\right)^{-8/11}, \quad (136)$$

$$\frac{P_r}{P_g} = 12 M_8^{2/11} L_*^{-2/11} (1-\beta)^{15/11} \xi_0^{-15/11} \left(\frac{r}{20}\right)^{6/11} \left(\frac{C_{bf}}{30}\right)^{6/11}, \quad (137)$$

Note that the $r$ dependences in equations (134)-(137) are quite different from those in equations (52)-(55). The differences result from the substitution of $(1 - \xi_0)$ from equation (131) into the latter equations. In particular, with $\xi_0$ given by equation (131), the ratio $P_r/P_g$ increases with $r$ since the dependence on $r$ through $\xi_0$ in equation (55) overcomes the explicit $r$ dependence in that equation. Equation (137) implies that $P_r > P_g$ for $r \gtrsim 6$, provided that equation (131) is satisfied. Equation (131) has a solution for

$$L_* \gtrsim \left(\frac{r}{20}\right)^{93/64} M_8^{-1/32} \left(\frac{C_{bf}}{30}\right)^{1/4}. \quad (138)$$

Thus for any given $L_*$ there is a maximal radial extent of the radiation pressure region

$$r \lesssim 20 \left(\frac{L_*}{0.1}\right)^{64/93} M_8^{2/93} \left(\frac{C_{bf}}{30}\right)^{-16/93}. \quad (139)$$

From equation (131) and (135) follows that (at fixed $r$) an increase of $L_*$ causes an increase in $\tau_{es}$. Therefore, the radiation pressure region is secularily stable. The physical explanation for this stability is rather simple. Equation (53) indicates that $\tau_{es}$ is inversely proportional to $L_*$ and to $(1 - \xi_0)^2$. From equation (131) follows that $(1 - \xi_0) \propto L_*^{-32/33}$, thus the resulting increase in $\tau_{es}$ overcomes the decrease due to increase of $L_*$.

The line marked R in figure 1 shows the luminosity versus the electron scattering optical depth of the disk (proportional to the disk surface density), in the gas pressure region for $r = 20$, $\beta = 0.2$, $M_8 = 1$ and $C_{bf} = 30$. The plot represents equation (135) with $\xi_0$ obtained from equation (131). The positive slope indicates that the solution is secularily stable, in contrast with the prediction of the $\alpha$ disk model.

**Discussion**

This paper presents a self-consistent solution for thin accretion disks with turbulent convection. The energy release is due to the interaction of turbulent convection with the Keplerian shear via a turbulent viscosity. Since turbulent convection provides the disk viscosity, the convective flux (in addition to the radiative flux) could be important in the transport of the generated energy to the disk surface. Employing a model for turbulence, the turbulent viscosity and the convective flux were obtained as functions of the physical parameters of the disk, which in turn are controlled by the former two. Having a model for turbulence, there is no need to resort to phenomenological parameterizations of either the viscosity (as done in the $\alpha$-disk models) or the convective flux (as done in the mixing length approach).

Solutions for both, radiation pressure dominated (inner) and for gas pressure dominated (outer) regions, with either electron scattering or free-free absorption opacities, were obtained. They provide the midplane temperature, density, disk half-width and the ratio between the convective and total fluxes at a given distance, for given central mass and given accretion rate. The solutions in the various regions of the disk depend on $\xi_0$— the fraction of the convective flux out of the total flux. We use an approximate estimate for the ratio of the surface to midplane temperatures as well as expressions for the flux emerging from the disk top, in terms of $T_s$. This yields an estimate for $\xi_0$ as function of the local physical conditions, in the different regions.

In the gas pressure dominated regions we find two distinct types of solutions. The first (corresponding to strong convection) is characterized by a convective flux much larger than the radiative flux. In this case, the blackbody region encompasses the entire gas pressure region and could as well extend down to the inner disk boundary, with no radiation pressure region. In the second class of solutions (corresponding to moderate convection), the convective flux is of the order of the radiative flux ($\sim 1/3$ of total flux) and all three disk-regions would exist. In the radiation pressure dominated region $\xi_0$ increases with $L_*$ and decreases with $r$. Thus, at $r = 20$, it ranges from $\sim 0.5$ when $L_* \sim 0.1$ up to very close to 1 when $L_* \gtrsim 1$. The radial extent of the radiation pressure region is larger the larger is the accretion rate. The radiation pressure region is optically thick and geometrically thin even for super Eddington luminosities.

All the disk solutions, including the radiation pressure dominated region, are secularily stable. In the radiation pressure region, an increase in the accretion rate causes an increase in the fraction of the convective flux. As a result the net dependence of the surface density on the accretion rate is through a positive power. The present solutions differ markedly from the $\alpha$ disk behavior, also in additional aspects. In the case of $\alpha$ disks, the radiation pressure dominated region becomes optically thin and geometrically thick for luminosities close to the Eddington luminosity. In the present case, all the solutions are geometrically thin and optically thick, even for luminosities equal to or exceeding the Eddington luminosity (see Fig. 1). Therefore, for luminosities of the order of the Eddington luminosity, the local spectrum will be a modified blackbody while in $\alpha$ disks it will be a bremsstrhalung or compotonized bremsstrhalung.

It is interesting to note that the trend of convection to result in larger surface density, lower temperature and increased stability, is evident even in models where the disk viscosity is described by an $\alpha$ parameter and the convective flux by the mixing length approach (Milsom *et al.* 1994)

We found that, for the same central mass and accretion rate, the innermost regions could be either radiation pressure dominated or gas pressure dominated blackbody with the convective flux much larger than the radiative flux. While the values of the corresponding effective $\alpha$-parameter differ only by a factor of $\sim 2$ (see below), in the former case the disk is much thicker geometrically, much less dense and hotter than in the second case. The emerging flux is a modified blackbody while in the second case it is a blackbody. Since most of the disk emission comes from the innermost regions, the spectral signature of these two solutions will be quite different for the same value of bolometric luminosity, corresponding to the same accretion rate.

As discussed above, the present disk model differs markedly from $\alpha$ disk models in various aspects. Nevertheless, it seems that the direct observational distinction between the two is not a straightforward one. For a given accretion rate, the temperatures and densities in the optically thick regions of an $\alpha$ disk are of a similar order of magnitude as in the present model, thus leading to similar spectra. As mentioned above, for high luminosities the spectra in the radiation pressure region would be quite different: a bremssrahlung or compotonized bremsstrhalung spectrum in the case of alpha disk versus a modified blackbody in the present case. However, an optically thin corona on top of an optically thick disk would also produce a hot bremsstrahlung or compotonized bremsstrhalung spectrum (Haardt & Maraschi 1991, 1993).

The values of the effective $\alpha$-parameter in the different regions are quite small: $\lesssim 5 \times 10^{-4}$, $\sim 1 \times 10^{-3}$, and $\sim 5 \times 10^{-3}$ for radiation pressure region and for the two types of solutions in the gas pressure region, respectively. These small values reflect the fact that the turbulent velocities are subsonic and that the horizontal scale of the largest eddies is small compared to the vertical scale. This anisotropy, resulting from the Coriolis force, is a general feature of a three-dimensional turbulence in a rotating disk. The above values are small compared to values required to model outburst of cataclysmic variables (see, e.g., Duschl 1989). This implies that turbulent convection cannot account for the disk viscosity in this case. A different mechanism, possibly magnetic viscosity, is required. One may speculate that turbulent convection can be the source of the disk viscosity during the quiescent state and some other mechanism provides the disk viscosity during the outburst.

Finally, we wish to stress the need for a model capable of providing a detailed vertical structure yet being selfconsistent. This is no easy task as present available spectral turbulence models, like the one applied here, assume homogeneity in the vertical direction and thus can yield only vertically averaged description. Also, the incorporation of the interaction of shear with the eddies is an open question in such models. A promising alternative is the Reynold stress approach that does not provide the turbulence spectral function but rather supplies differential equations in space and time for the various turbulence ensemble averages. The formalism has been employed, extensively and quite successfully, in atmospheric and laboratory turbulence (Zeman 1981; Speziale 1991) and has been used recently to study convective overshooting (Canuto 1992, 1993) in stars.

This work was supported by the Tel Aviv University Research Fund, grant 405-93.

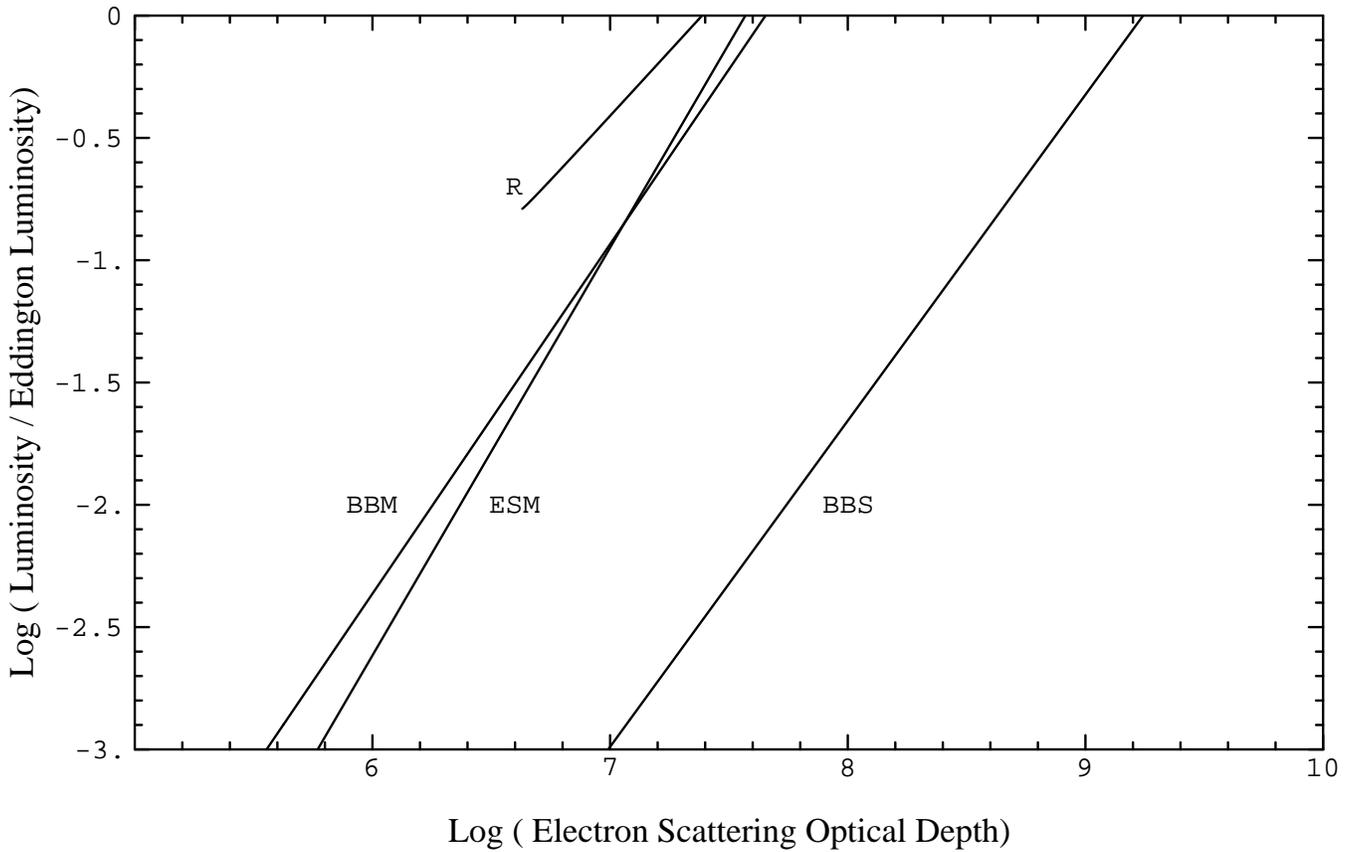

Fig. 1.—Lminosity, in units of the Eddington luminosity vs. the electron-scattering optical depth, for the various convective disk solutions, at a constant distance $r = 20$, for $M_8 = 1$ and $C_{bf} = 30$. $R$: radiation pressure region, $\beta = 0.2$, eqs. (135), (131). $BBS$: gas pressure blackbody solution in the strong convection limit, eq. (93). $BBM$: same region for the case of moderate convection, eq. (109). $ESM$: gas pressure electron-scattering solution in the moderate convection limit, eq. (117).